\documentclass[journal]{IEEEtran}

\usepackage{cite}
\usepackage{epsfig}
\usepackage{epstopdf}
\usepackage{graphicx}
\usepackage{amsfonts}
\usepackage{amsmath,bm}
\usepackage{amssymb}
\usepackage{multicol}
\usepackage{psfrag}
\usepackage[caption=false]{subfig}
\usepackage{stfloats}
\usepackage{footnote}
\usepackage{array}
\usepackage{algorithm}
\usepackage{algorithmic}
\usepackage{tikz}
\usepackage{mathtools}
\usepackage{multirow}

\DeclareMathOperator*{\argmax}{argmax}
\DeclareMathOperator*{\argmin}{argmin}

\allowdisplaybreaks

\begin{document}
\title{6D Movable Antenna Enhanced Interference Mitigation for Cellular-Connected UAV Communications}

\author{Tianshi Ren, Xianchao Zhang,~\IEEEmembership{Member,~IEEE,} Lipeng Zhu,~\IEEEmembership{Member,~IEEE,} Wenyan Ma,~\IEEEmembership{Graduate Student Member,~IEEE,} Xiaozheng Gao,~\IEEEmembership{Member,~IEEE,} and Rui Zhang,~\IEEEmembership{Fellow,~IEEE}%
\thanks{T. Ren is with the School of Information and Electronics, Beijing Institute of Technology, Beijing 100081, China, and also with the Department of Electrical and Computer Engineering, National University of Singapore, Singapore 117583 (email: rentianshi@bit.edu.cn).}
\thanks{X. Zhang is with the Provincial Key Laboratory of Multimodal Perceiving and Intelligent Systems, and also with the Engineering Research Center of Intelligent Human Health Situation Awareness of Zhejiang Province, Jiaxing University, Jiaxing 314001, China (e-mail: zhangxianchao@zjxu.edu.cn).}
\thanks{L. Zhu and W. Ma are with the Department of Electrical and Computer Engineering, National University of Singapore, Singapore 117583 (email: zhulp@nus.edu.sg, wenyan@u.nus.edu).}
\thanks{X. Gao is with the School of Information and Electronics, Beijing Institute of Technology, Beijing 100081, China (email: gaoxiaozheng@bit.edu.cn).}
\thanks{R. Zhang is with School of Science and Engineering, Shenzhen Research Institute of Big Data, The Chinese University of Hong Kong, Shenzhen, Guangdong 518172, China (e-mail: rzhang@cuhk.edu.cn). He is also with the Department of Electrical and Computer Engineering, National University of Singapore, Singapore 117583 (e-mail: elezhang@nus.edu.sg).}}

\maketitle

\begin{abstract}

Cellular-connected unmanned aerial vehicle (UAV) communications is an enabling technology to transmit control signaling or payload data for UAVs through cellular networks. Due to the line-of-sight (LoS) dominant air-to-ground channels, efficient interference mitigation is crucial to UAV communications, while the conventional fixed-position antenna (FPA) arrays have limited degrees of freedom (DoFs) to suppress the interference between the UAV and its non-associated co-channel base stations (BSs). To address this challenge, we propose in this letter a new approach by utilizing the six-dimensional movable antenna (6DMA) arrays to enhance the interference mitigation for the UAV. Specifically, we propose an efficient block coordinate descent (BCD) algorithm to iteratively optimize the antenna position vector (APV), array rotation vector (ARV), receive beamforming vector, and associated BS of the UAV to maximize its signal-to-interference-plus-noise ratio (SINR). Numerical results show that the proposed 6DMA enhanced cellular-connected UAV communication can significantly outperform that with the traditional FPA arrays and other benchmark schemes in terms of interference mitigation.

\end{abstract}


\IEEEpeerreviewmaketitle

\section{Introduction}\label{sec:intro}

In the last decade, integrating the unmanned aerial vehicles (UAVs) into wireless communication networks has emerged as a promising paradigm by utilizing the deployment of low-cost and on-demand aerial nodes with the communication capability ~\cite{Zeng2019Accessing}. On one hand, UAVs can be deployed as the aerial base stations (BSs) or relays to support the communication of terrestrial users. On the other hand, UAVs can be integrated into the cellular network as new aerial users. In the latter scenario (i.e., cellular-connected UAVs), UAVs have a larger operation range compared to terrestrial users, enabling them to establish communication links with more ground BSs, while this also leads to increased interference with more non-associated  BSs operating over the same frequency band ~\cite{Fotouhi2019Survey,Mei2020Cooperative}. To address this problem, multi-antenna technologies are usually adopted at the UAV to suppress the interference via beamforming~\cite{Zeng2019Accessing,Fotouhi2019Survey}. However, since the conventional UAVs are equipped with fixed-position antenna (FPA) arrays of a fixed geometry, the steering vectors of the array over different steering angles are fixed. Therefore, existing interference mitigation methods with FPA arrays cannot change the inherent correlation between the steering vectors over varying associated BS's and non-associated BSs' directions. Instead, only the beamforming vector of the UAV can be designed for interference mitigation, which limits the degrees of freedom (DoFs) in interference mitigation to adapt to the time-varying channels between the UAV and non-associated BSs.

Recently, movable antenna (MA) has been recognized as a promising technology for efficiently improving the wireless channel conditions and enhancing the communication performance via the local movement of antennas~\cite{Zhu2023Movable}, which is also known as fluid antenna system~\cite{Zhu2024Historical}. Preliminary studies have demonstrated the advantages of MAs in improving the received signal-to-noise ratio (SNR)~\cite{Zhu2023Movable}, the multi-input-multi-output (MIMO) channel capacity~\cite{Ma2023MIMO,Ye2024Fluid}, the performance of multiuser communications~\cite{Zhu2023Modeling}, and flexible beamforming~\cite{Ma2024Multi,Zhu2023Movablea} over their FPA counterparts. Moreover, it has been shown in~\cite{Liu2024UAV} and~\cite{Kuang2024Movable} that an MA array can significantly enhance the UAV-terrestrial user communication, where the UAV serves as an aerial BS. In~\cite{Liu2024UAV}, the achievable data rate was maximized in UAV-enabled multi-user multi-input-single-output (MISO) systems enhanced by the MA array with joint optimization of transmit beamforming, the UAV trajectory, and the positions of MAs. In~\cite{Kuang2024Movable}, the data rate was maximized in an integrated sensing and communications system deployed in the UAV-enabled low-altitude platform aided by the MA array via joint optimization of the transmit information beamforming, sensing beamforming, and the MAs' positions. In addition, the six-dimensional MA (6DMA) was proposed in~\cite{Shao20246D,Shao20246DMA} to improve multiuser communication performance by jointly optimizing the antenna position and rotation at the ground BS. However, to the best of our knowledge, no prior work has investigated MA enhanced cellular-connected UAV communications. 

In this letter, we propose utilizing the 6DMA array to enhance the interference mitigation caused by the co-channel terrestrial transmissions to the UAV. Our objective is to maximize the received signal-to-interference-plus-noise ratio (SINR) of the UAV by jointly optimizing its antenna position vector (APV), array rotation vector (ARV), receive beamforming vector, and selection of the associated BS. To handle this non-convex optimization problem, we first express the optimal receive beamforming vector as a function of other variables. Then, we introduce an auxiliary vector to address the minimum inter-antenna distance constraint. Subsequently, we employ the block coordinate descent (BCD) method to alternately optimize the APV, auxiliary vector, and ARV for each selection of the associated BS. Numerical results demonstrate that the proposed 6DMA enhanced cellular-connected UAV communication scheme can significantly outperform that with the traditional FPA arrays and other benchmark schemes in terms of interference mitigation.

\textit{Notations:} $x$, $\mathbf{x}$, $\mathbf{X}$ denote a scalar, vector, and matrix, respectively. $(\cdot)^T$, $(\cdot)^H$, and $(\cdot)^{-1}$ denote transpose, conjugate transpose and inverse, respectively. $\Vert\mathbf{x}\Vert$ denotes the 2-norm of the vector $\mathbf{x}$. $\mathbb{R}$ and $\mathbb{C}$ denote the set of real numbers and complex numbers, respectively. $\mathcal{A}\cup\mathcal{B}$ denotes the union of set $\mathcal{A}$ and set $\mathcal{B}$, and $\mathcal{A}\backslash\mathcal{B}$ denotes the set of elements that belong to $\mathcal{A}$ but are not in $\mathcal{B}$. $\lceil x\rceil$ denotes the ceil of a real number $x$.

\section{System Model and Problem Formulation}\label{sec:model}

\subsection{System Model}\label{subsec:system}
\begin{figure}
		\centering
		\includegraphics[width=80mm]{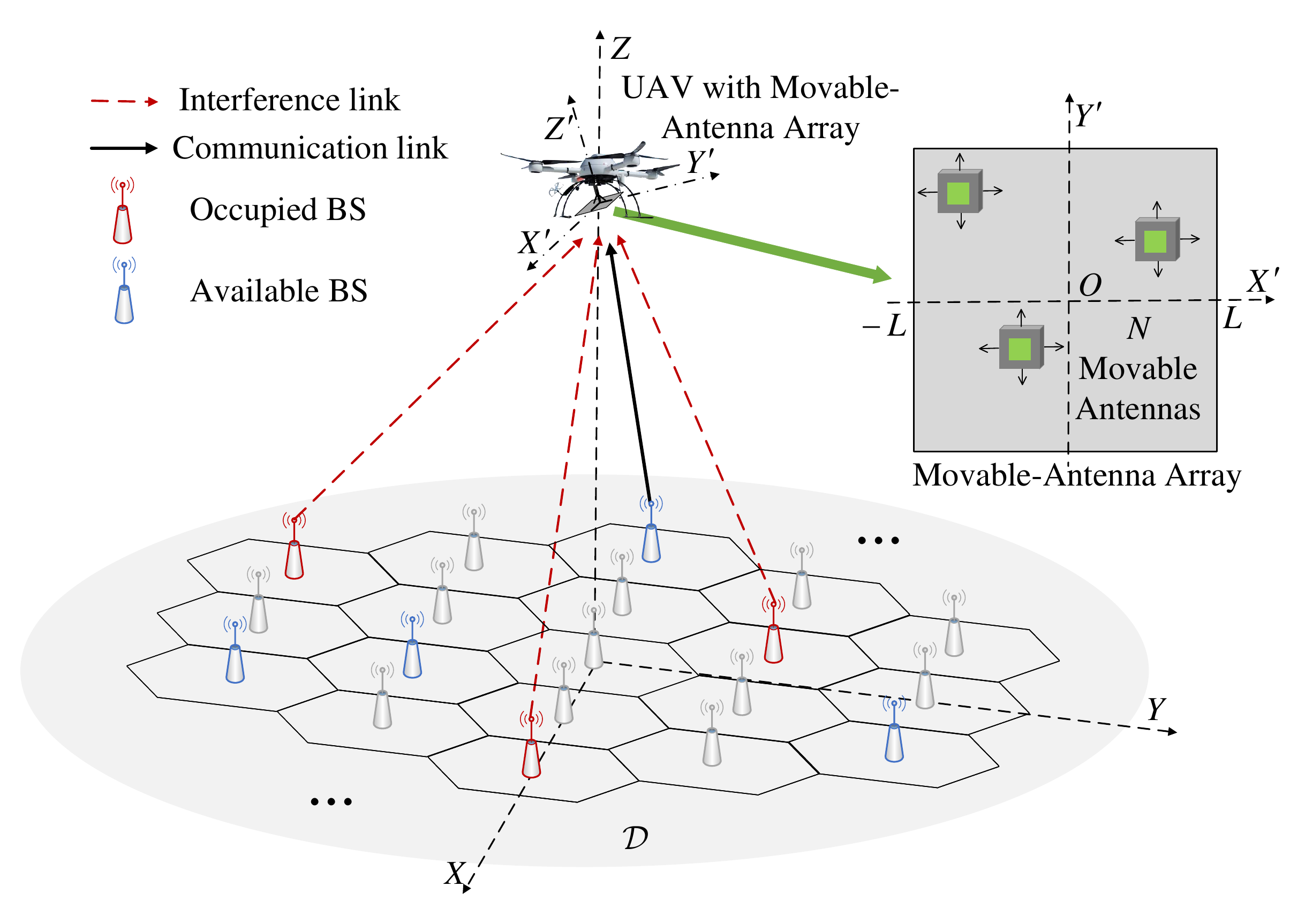}
            \caption{The downlink cellular-connected UAV communication system.}
            \label{fig:model1}
            \vspace{-0.4cm} 
\end{figure}

As shown in Fig.~\ref{fig:model1}, we consider a downlink cellular-connected UAV communication system, with $M$ BSs in a given interfering zone $\mathcal{D}$. In the case of hexagon cell shape, the $M$ BSs form the $q$-tier neighboring BSs (including the nearest BS) of the BS nearest to the UAV. BSs outside $\mathcal{D}$ are unable to establish communication links with the UAV due to excessive path loss, and thus do not contribute to interference in the considered UAV communication system. Due to the reuse of time-frequency resources, some terrestrial users may utilize the same resource block (RB) as the UAV, leading to co-channel interference. Let $J$ represent the number of terrestrial users within $\mathcal{D}$ which share the same RB with the UAV. Since one BS can serve only one user per RB, we define the BS serving its associated co-channel terrestrial user as the \textit{occupied BS}. Accordingly, the number of occupied BSs is $J$, where the set of all occupied BSs is denoted as $\mathcal{J}$. 

To mitigate inter-cell interference (i.e., interference between users sharing the same RB), we introduce the concept of the inter-cell interference coordination (ICIC)~\cite{Mei2020Cooperative}. Specifically, the first $e$-tier neighboring BSs of each occupied BS are prohibited from serving the terrestrial users or UAV on the same RB. We define the BSs that remain unoccupied and meet the ICIC condition as \textit{available BSs}, which can be selected to establish communication links with the UAV. Let $\mathcal{T}_a(b)$ denote the set of the first $b$-tier neighboring BSs of BS $a$ including itself. The set of the available BSs is then expressed as
\begin{align}
\mathcal{K}=\mathcal{T}_c(q)\backslash\{\bigcup_{i\in\mathcal{J}}\mathcal{T}_i(e)\},
\end{align}
where BS~$c$ is the BS nearest to the UAV, and $K\triangleq |\mathcal{K}|$. 

As shown in Fig.~\ref{fig:model1}, the UAV is equipped with $N$ MAs, each of which can move within a two-dimensional (2D) square panel of side length $2L$. Moreover, the panel is a 6DMA array that can rotate in three dimensions to reconfigure the angles of arrival (AoAs)  of ground-to-air line-of-sight (LoS) channels. In the three-dimensional (3D) ground coordinate system $X$-$Y$-$Z$, the position of BS~$m$ is given by $\mathbf{p}_m=\left[X_m,Y_m,Z_m\right]^T, 1\leq m\leq M$, and the position of UAV is denoted by $\mathbf{p}_u=\left[X_u,Y_u,Z_u\right]^T$. The vector from BS~$m$ to the UAV can be expressed as $\tilde{\mathbf{v}}_{m}=\mathbf{p}_u-\mathbf{p}_m$. To characterize the rotation of 6DMA array, we consider the local 3D coordinate system $X'$-$Y'$-$Z'$ with its origin being the center of the 6DMA array. The ARV is defined as $\mathbf{a} = [\phi, \psi, \theta]^T$, where $\phi$, $\psi$, and $\theta$ denote the array rotation angles with respect to (w.r.t.) the $X'$-axis, $Y'$-axis and $Z'$-axis, respectively. Then, the 6DMA array rotation matrices are given by 
\begin{align}
\mathbf{R}_x(\phi)=\begin{bmatrix}
1 & 0 & 0 \\
0 & \cos{\phi} & -\sin{\phi} \\
0 & \sin{\phi} & \cos{\phi}
\end{bmatrix},
\end{align}
\begin{align}
\mathbf{R}_y(\psi)=\begin{bmatrix}
\cos{\psi} & 0 & \sin{\psi} \\
0 & 1 & 0 \\
-\sin{\psi} & 0 & \cos{\psi}
\end{bmatrix},
\end{align}
\begin{align}
\mathbf{R}_z(\theta)=\begin{bmatrix}
\cos{\theta} & -\sin{\theta} & 0 \\
\sin{\theta} & \cos{\theta} & 0 \\
0 & 0 & 1
\end{bmatrix}.
\end{align}
Consequently, the overall rotation matrix of the 6DMA array can be expressed as
\begin{align}
\mathbf{U}(\mathbf{a})=\mathbf{R}_x(\phi)\mathbf{R}_y(\psi)\mathbf{R}_z(\theta).
\end{align}

With the rotation of the 6DMA array, in the local 3D coordinate system $X'$-$Y'$-$Z'$, the vector from BS~$m$ to the UAV is given by $\breve{\mathbf{v}}_{m}=\mathbf{U}(\mathbf{a})^T\tilde{\mathbf{v}}_{m}$. Therefore, the wave vector from BS~$m$ to the UAV is given by
\begin{align}
\mathbf{v}_{m}=\cfrac{\breve{\mathbf{v}}_{m}}{\Vert\breve{\mathbf{v}}_{m}\Vert}\triangleq[\alpha_{m},\beta_{m},\delta_{m}]^T.
\end{align}

In the local 2D MA panel coordinate system $X'$-$Y'$, the position of MA~$n$ is denoted by $\mathbf{x}_{n}=[x_{n},y_{n}]^T$. The APV of all MAs is then represented as $\tilde{\mathbf{x}}=[\mathbf{x}_{1}^T,\mathbf{x}_{2}^T,\ldots,\mathbf{x}_{N}^T]^T\in\mathbb{R}^{2N}$. Thus, the array-response vector for the LoS channel between the UAV and BS~$m$ can be expressed as 
\begin{align}
\mathbf{g}_{m}=\left[e^{j\frac{2\pi}{\lambda}\rho_{m,1}},e^{j\frac{2\pi}{\lambda}\rho_{m,2}},\dots,e^{j\frac{2\pi}{\lambda}\rho_{m,N}}\right]^T,
\end{align}
where $\lambda$ is the carrier wavelength, and $\rho_{m,n} = x_{n}\alpha_{m}+y_{n}\beta_{m}$ represents the difference in signal propagation distance between MA~$n$ at position $\mathbf{x}_{n}$ and the origin $O$ of the local 2D MA panel coordinate system $X'$-$Y'$.

Given that the ground-to-air channel between the UAV and the ground BS is dominated by the LoS link, the channel vector between the UAV and the BS~$m$ is expressed as
\begin{align}
\mathbf{h}_{m}=\cfrac{\lambda}{4\pi d_m}e^{-j\frac{2\pi}{\lambda}d_{m}}\mathbf{g}_{m},
\end{align}
where $d_{m}=\Vert\mathbf{p}_{u}-\mathbf{p}_m\Vert$ is the distance between the UAV and BS~$m$.
            
When the UAV selects one of the available BS (i.e., BS $k$, $k\in\mathcal{K}$) as the associated BS, the received signal at the UAV is given by
\begin{align}
y_{k}=\mathbf{w}^H\mathbf{h}_{k}\sqrt{P}s_{u}+\mathbf{w}^H\sum\limits_{i\in \mathcal{J}}{\mathbf{h}_{i}\sqrt{P}s_{i}}+\mathbf{w}^H\mathbf{z},~k\in\mathcal{K},
\end{align}
where $\mathbf{w}\in \mathbb{C}^N$ is the receive beamforming vector of the UAV; $P$ denotes the transmit power of each BS over the considered RB; $s_{u}$ and $s_{i}$ are normalized symbols for the UAV and the terrestrial users served by BS $i$, respectively; and $\mathbf{z}\sim\mathcal{CN}(\mathbf{0},\sigma^2\mathbf{I}_N)$ represents the additive white Gaussian noise (AWGN) at the UAV with average power $\sigma^2$. Thus, the received SINR at the UAV can be expressed as
\begin{align}
\gamma_{k}=\cfrac{|\mathbf{w}^H\mathbf{h}_{k}|^2P}{\sum\limits_{i\in\mathcal{J}}|\mathbf{w}^H\mathbf{h}_{i}|^2P+\Vert\mathbf{w}\Vert^2\sigma^2},~k\in\mathcal{K}.
\end{align}	                    

\subsection{Problem Formulation}\label{subsec:problem}
In this letter, we aim to maximize the received SINR of the UAV by jointly optimizing its APV, ARV, receive beamforming vector, and selection of the associated BS. Accordingly, the SINR maximization problem can be formulated as
\begin{subequations}
		\begin{align}
                \textrm {(P1)}~~\max\limits_{k\in \mathcal{K}}&\max\limits_{\mathbf{w},\tilde{\mathbf{x}},\mathbf{a}}~\gamma_{k} \label{P1a}\\
			\text{s.t.} \quad & -L \leq x_n \leq L,\forall n,\label{P1b}& \\ 
               & -L \leq y_n \leq L,\forall n,\label{P1c}& \\
			 &\Vert \mathbf{x}_m-\mathbf{x}_n\Vert\geq L_0,~\forall m,n, m\neq n,\label{P1d}&
		\end{align}
\end{subequations}
where $L_0$ in constraint~(\ref{P1d}) is the minimum inter-antenna distance to avoid coupling. Constraints~(\ref{P1b}) and~(\ref{P1c}) ensure that the movement of MAs is confined within the specified range. The optimization problem (P1) is challenging to be solved optimally due to the non-convexity introduced by the APV and ARV. Consequently, in the next section, we propose a low-complexity algorithm to obtain suboptimal solutions for problem (P1).

\section{Proposed Solution}\label{sec:solution}

 In this section, we first express the optimal beamforming vector as the function of the other variables in (P1). Then, we introduce an auxiliary vector and employ the penalty method to deal with the minimum inter-antenna distance constraint. Finally, we propose a BCD-based algorithm to alternatively optimize the APV, the auxiliary vector, and the ARV for each available BS. After obtaining the UAV-received SINR for all available BSs, we select the available BS achieving the highest SINR as the associated BS.

\subsection{Beamforming Vector Optimization}\label{subsec:W}

When $\mathbf{a}$, $\tilde{\mathbf{x}}$ and $k$ are fixed, an optimal receive beamforming vector is obtained by the minimum mean square error (MMSE) method as \cite{Tse2005Fundamentals} 
 \begin{align}
 \mathbf{w}^*=\left(\mathbf{I}_N+\sum\limits_{i\in\{\mathcal{J},k\}}\frac{P}{\sigma^2}\mathbf{h}_i\mathbf{h}_i^H\right)^{-1}\mathbf{h}_k.
 \end{align}
 Therefore, the optimal receive beamforming vector can be expressed as the function of the APV, ARV, and the index of the associated BS. Substituting $\mathbf{w}^*$ into the original problem (P1), the optimization of the APV and ARV can be decomposed into two subproblems, which are respectively solved in the following.

\subsection{APV Optimization}\label{subsec:X}

For any given $\mathbf{a}$ and $k$, (P1) can be reformulated as 
 \begin{subequations}
		\begin{align}
                \textrm {(P2)}~\max\limits_{\tilde{\mathbf{x}}} &~\gamma_{k}=\cfrac{|\mathbf{h}_k^H\mathbf{B}^H\mathbf{h}_k|^2P}{\sum\limits_{i\in\mathcal{J}}|\mathbf{h}_k^H\mathbf{B}^H\mathbf{h}_i|^2P+\Vert\mathbf{B}\mathbf{h}_k\Vert^2\sigma^2} \label{P3a}\\
			\text{s.t.} \quad 
            & -L \leq x_n \leq L,\forall n,\label{P3b}& \\ 
               & -L \leq y_n \leq L,\forall n,\label{P3c}& \\
			 &\Vert \mathbf{x}_m-\mathbf{x}_n\Vert\geq L_0,~\forall m, n, m\neq n,\label{P3d} & 
		\end{align}
\end{subequations}
where $\mathbf{B}\triangleq\left(\mathbf{I}_N+\sum\limits_{i\in\{\mathcal{J},k\}}\frac{P}{\sigma^2}\mathbf{h}_i\mathbf{h}_i^H\right)^{-1}$.

Problem (P2) is still difficult to solve since the objective function~(\ref{P3a}) is non-concave and the constraint~(\ref{P3d}) is non-convex. Therefore, we introduce an auxiliary vector $\tilde{\mathbf{r}}=[\mathbf{r}_{1}^T,\mathbf{r}_{2}^T,\ldots,\mathbf{r}_{N}^T]^T\in\mathbb{R}^{2N}$ and incorporate a penalty term into the objective function. The penalty function $p(\tilde{\mathbf{x}},\tilde{\mathbf{r}})$ is expressed as
 \begin{align}
                p(\tilde{\mathbf{x}},\tilde{\mathbf{r}})  =\mu\sum\limits_{n=1}^{N}\Vert\mathbf{x}_n-\mathbf{r}_n\Vert^2,
\end{align}
where $\mu$ is the penalty factor. Thus, (P2) is transferred into the following optimization problem:			
 \begin{subequations}
		\begin{align}
                \textrm {(P3)}~\min\limits_{\tilde{\mathbf{x}},\tilde{\mathbf{r}}} &~-\cfrac{|\mathbf{h}_k^H\mathbf{B}^H\mathbf{h}_k|^2P}{\sum\limits_{i\in\mathcal{J}}|\mathbf{h}_k^H\mathbf{B}^H\mathbf{h}_i|^2P+\Vert\mathbf{B}\mathbf{h}_k\Vert^2\sigma^2} + p(\tilde{\mathbf{x}},\tilde{\mathbf{r}})  \label{P4a}\\
			\text{s.t.} \quad 
            & -L \leq x_n \leq L,\forall n,\label{P4b}& \\ 
               & -L \leq y_n \leq L,\forall n,\label{P4c}& \\
			 &\Vert \mathbf{r}_m-\mathbf{r}_n\Vert\geq L_0,~\forall m, n, m\neq n. \label{P4d}& 
		\end{align}
\end{subequations}
If $\tilde{\mathbf{x}}=\tilde{\mathbf{r}}$, the penalty function equals $0$; otherwise, the objective function incurs a penalty during the iteration of the algorithm.

Given that constraints (\ref{P4b}) and (\ref{P4c}) pertain only to $\tilde{\mathbf{x}}$ and (\ref{P4d}) pertains only to $\tilde{\mathbf{r}}$, (P3) can be decomposed into two subproblems, namely (P3-a) and (P3-b) as follows. 
 
 \textit{1) Optimization of APV $\tilde{\mathbf{x}}$}
 
 For fixed $\tilde{\mathbf{r}}$, (P3) can be expressed as
 \begin{subequations}
		\begin{align}
                \textrm {(P3-a)}~\min\limits_{\tilde{\mathbf{x}}} &~f(\tilde{\mathbf{x}})\triangleq-\cfrac{|\mathbf{h}_k^H\mathbf{B}^H\mathbf{h}_k|^2P}{\sum\limits_{i\in\mathcal{J}}|\mathbf{h}_k^H\mathbf{B}^H\mathbf{h}_i|^2P+\Vert\mathbf{B}\mathbf{h}_k\Vert^2\sigma^2}\nonumber\\& + p(\tilde{\mathbf{x}},\tilde{\mathbf{r}})\label{P4aa}\\
			\text{s.t.} \quad 
            & -L \leq x_n \leq L,\forall n,\label{P4ab}& \\ 
               & -L \leq y_n \leq L,\forall n.\label{P4ac}& 
               \end{align}
\end{subequations}

Since (P3-a) only imposes constraints on the moving region, the projected gradient descent (PGD) algorithm can be employed to optimize the APV $\tilde{\mathbf{x}}$. Specifically, let $\tilde{\mathbf{x}}^i$ represent the APV in the $i$-th iteration of the PGD algorithm. Then, the gradient of the objective function is calculated as
\begin{align}
       \left[\triangledown f(\tilde{\mathbf{x}}^i)\right]_n=\lim\limits_{\epsilon\rightarrow0}\cfrac{f(\tilde{\mathbf{x}}^i+\epsilon\mathbf{e}_n)}{\epsilon},~n=1,\ldots,2N,    
\end{align}
where $\mathbf{e}_n$ is a $2N$-dimensional vector with its $n$-th element being $1$ and the other elements being $0$.

We can take the opposite direction of the gradient as the descent direction $\mathbf{d}^i=-\triangledown f(\tilde{\mathbf{x}}^i)$. Then, the APV $\tilde{\mathbf{x}}^i$ can be updated as
\begin{align}
      \tilde{\mathbf{x}}^{i+1} = \mathcal{P}\left(\tilde{\mathbf{x}}^{i}+\alpha^i\mathbf{d}^i\right),
\end{align}
where $\alpha^i$ is the step size of $i$-th iteration, and the element-wise projection function $\mathcal{P}(x)$ can be expressed as
\begin{align}
      \mathcal{P}\left(x\right) = \begin{cases}
-L& \text{if } x\leq-L,\\
x& \text{if } -L<x<L,\\
L& \text{if } x\geq L.
\end{cases}
\end{align}

In addition, we use the backtracking method to get the appropriate step size that satisfies the Armijo–Goldstein condition:
\begin{align}
       f(\tilde{\mathbf{x}}^{i+1})\leq f(\tilde{\mathbf{x}}^i)+c\alpha^i\triangledown f(\tilde{\mathbf{x}}^i)^T\mathbf{d}^i,     
\end{align}
where $c\in[0,1]$ is a constant to achieve an adequate decrease in the objective function with step size $\alpha^i$.

The above PGD algorithm terminates when the increase of (\ref{P4aa}) is below a predefined threshold, thus we can obtain a suboptimal APV solution $\tilde{\mathbf{x}}$ for (P3-a). 

\textit{2) Optimization of Auxiliary Vector $\tilde{\mathbf{r}}$}

For fixed $\tilde{\mathbf{x}}$, (P3) can be expressed as
\begin{subequations}
		\begin{align}
                \textrm {(P3-b)}~\min\limits_{\tilde{\mathbf{r}}} &~p(\tilde{\mathbf{r}})=\sum\limits_{n=1}^{N}\Vert\mathbf{x}_n-\mathbf{r}_n\Vert^2 \label{P4ba}\\
			\text{s.t.} \quad 
			 &\Vert \mathbf{r}_m-\mathbf{r}_n\Vert\geq L_0,~\forall m,n, m\neq n\label{P4bb}. & 
		\end{align}
\end{subequations}

To solve the non-convex optimization problem (P3-b), we decouple it into $N$ subproblems and solve them iteratively, where each subproblem optimizes one variable in $\{\mathbf{r}_n\}_{n=1}^N$ with the others being fixed. The optimization problem w.r.t. $\mathbf{r}_n$ can be expressed as
 \begin{subequations}
		\begin{align}
                \textrm {(P4)}~\min\limits_{\mathbf{r}_n} &~\Vert\mathbf{x}_n-\mathbf{r}_n\Vert^2 \label{P5a}\\
			\text{s.t.} \quad 
			 &\Vert \mathbf{r}_m-\mathbf{r}_n\Vert\geq L_0,~\forall m, m\neq n\label{P5b}. & 
		\end{align}
\end{subequations}

Although (P4) is a non-convex optimization problem due to the non-convex constraint~(\ref{P5b}), we can obtain the optimal solution of (P4) by considering the following two conditions~\cite{Jin2024Handling}:
   \begin{itemize}   
   \item[1.] If $\mathbf{x}_n$ satisfies the constraint $\Vert \mathbf{r}_m-\mathbf{x}_n\Vert\geq L_0,\forall m, m\neq n$, the optimal solution of (P4) is $\mathbf{r}_n^*=\mathbf{x}_n$.
    
    \item[2.] If $\mathbf{x}_n$ does not satisfy the constraint $\Vert \mathbf{r}_m-\mathbf{x}_n\Vert\geq L_0,~\forall m, m\neq n$, the set of all $\mathbf{r}_m$ satisfying $\Vert\mathbf{r}_m-\mathbf{x}_n\Vert < L_0$ is denoted as $\mathcal{R}$, and the circle centered on $\mathbf{r}_m$ with the radius $L_0$ is denoted as $\mathcal{C}_m$. For each $\mathbf{r}_j\in\mathcal{R}$, we check if there exist points on $\mathcal{C}_j$ satisfying~(\ref{P5b}) and collect the one nearest to $\mathbf{x}_n$ into set $\mathcal{S}$. Then, the optimal solution of (P4) is $\mathbf{r}_n^*=\argmin_{\mathbf{r}_n\in\mathcal{S}}\Vert\mathbf{x}_n-\mathbf{r}_n\Vert^2$. 
    
    \end{itemize}

\subsection{ARV Optimization}\label{subsec:A}

For fixed $\tilde{\mathbf{x}}$ and $k$, the objective function in (P1) can be rewritten as
		\begin{align}
                \textrm {(P5)}~\max\limits_{\mathbf{a}} &~\gamma_{k}=\cfrac{|\mathbf{h}_k^H\mathbf{B}^H\mathbf{h}_k|^2P}{\sum\limits_{i\in\mathcal{J}}|\mathbf{h}_k^H\mathbf{B}^H\mathbf{h}_i|^2P+\Vert\mathbf{B}\mathbf{h}_k\Vert^2\sigma^2} \label{P6a}.
		\end{align}
Since (P5) is an unconstrained optimization problem with a non-convex objective function, the gradient descent method with the similar procedure for solving (P3-a) can be employed to obtain a suboptimal solution for (P5). 

\begin{algorithm}[t!]
\caption{The BCD-based algorithm for solving (P1)}\label{BCD}
\begin{algorithmic}[1]
\REQUIRE$J,q,e,M,N, L,\sigma^2,L_0,\lambda,P,\mathbf{p}_u,\left\{\mathbf{p}_m\right\}_{m=1}^M$.
\ENSURE$\tilde{\mathbf{x}},\mathbf{a},\mathbf{w}^*,k^*$.
\FOR {$k = 1 : K$}
\STATE {Initialize }$\tilde{\mathbf{x}}$, $\tilde{\mathbf{r}}$, $\mathbf{a}$.
\REPEAT
\STATE{Calculate the APV }$\tilde{\mathbf{x}}${ by solving (P3-a)}.
\REPEAT
\FOR {$n = 1 : N$}
\STATE{Calculate 
 }$\mathbf{r}^*_n${ by solving (P4)}.
\ENDFOR
\UNTIL{The objective function of (P3-b) converges.}
\STATE{Calculate the ARV }$\mathbf{a}${ by solving (P5)}.
\UNTIL {Increase of $\gamma_k$ is below a predefined threshold.}
\STATE {Calculate the beamforming vector $\mathbf{w}^*$ according to (12).}
\ENDFOR
\STATE {$k^*\leftarrow\argmax_k\{\gamma_{k}\}_{k\in\mathcal{K}}$}
\end{algorithmic}
\label{alg_1}
\end{algorithm}

The overall BCD-based algorithm for solving (P1) is summarized in Algorithm~1. Specifically, we initialize the APV $\tilde{\mathbf{x}}$ by adopting the geometry of the uniform planar array (UPA) with $2L/(\lceil \sqrt{N} \rceil-1)$ inter-antenna spacing. The auxiliary vector is initialized as $\tilde{\mathbf{r}}=\tilde{\mathbf{x}}$, and the ARV is initialized as $\mathbf{a}=[0,0,0]^T$. The APV is optimized in line 4 using the PGD algorithm, the auxiliary variables $\{\textbf{r}_n\}_{n=1}^N$ are optimized in lines 5-9, and then the ARV is optimized in line 10 using the gradient descent algorithm. The algorithm terminates if the increase of the objective function is below a predefined threshold.

Moreover, the computation complexity is analyzed as follows. Let $T_x$ and $T_1$ represent the maximum number of iterations for solving (P3-a) and for the backtracking step search, respectively. Then, the total computation complexity of line 4 is $\mathcal{O}(T_xN+T_xT_1N)$. Similarly, the computation complexity of line 10 is $\mathcal{O}(T_aJ+T_aT_2J)$, where $T_a$ and $T_2$ denote the maximum number of iterations for solving (P5) and for the backtracking step search, respectively. Let $I_r$ and $I_k$ denote the maximum number of iterations for repeating lines 5-9 and lines 3-11, respectively. The total algorithm complexity is therefore in the order of $\mathcal{O}(KI_k(T_xN+T_xT_1N+I_rN^3+T_aJ+T_aT_2J))$.

\section{Numerical Results}\label{sec:results}

In this section, numerical results are provided to evaluate the performance of our proposed interference mitigation scheme for the 6DMA-enhanced cellular-connected UAV. We set the carrier wavelength as $\lambda=0.03$~m, and the minimum inter-antenna distance as $L_0=0.5\lambda$, i.e., $0.015$~m. The transmit power of each BS is $P=30 $~dBm, and the noise power is $\sigma^2=-109$~dBm. In the 3D ground coordinate system, the altitude of the UAV is $Z_u=100$~m, and the height of BS is $Z_m=30$~m,~$\forall m$. For the cellular network, the radius of each cell is $100$~m, the ICIC cell tier is set as $e=1$, and the number of UAV interfering cell tier is set as $q=4$, which means that the number of BSs in the interfering zone of the UAV is $M=1+3q(q+1)=61$. Furthermore, we compare our proposed method with the following baseline schemes. \textbf{Scheme 1} : The UAV selects its nearest BS as the associated BS, while the APV and ARV are iteratively optimized according to Algorithm 1. \textbf{Scheme 2} : The 6DMA array is always parallels to the $X$-$Y$ plane of the ground coordinate system with fixed ARV, while the APV is optimized according to Algorithm 1, and then the BS with highest SINR is selected as the associated BS. \textbf{Scheme 3} : The UAV selects its nearest BS as the associated BS, and the 6DMA array is always parallels to the $X$-$Y$ plane of the ground coordinate system, while the APV is optimized according to Algorithm 1. We also consider the FPA (UPA with $\lambda/2$ inter-antenna spacing) counterpart for each baseline scheme, where the MA array is replaced by the FPA array for comparison. All numerical results are averaged over 200 randomly generated positions of co-channel terrestrial users.

\begin{figure*}[htbp]
	\centering
	\begin{minipage}{0.32\linewidth}
		\centering
		\includegraphics[width=0.85\linewidth]{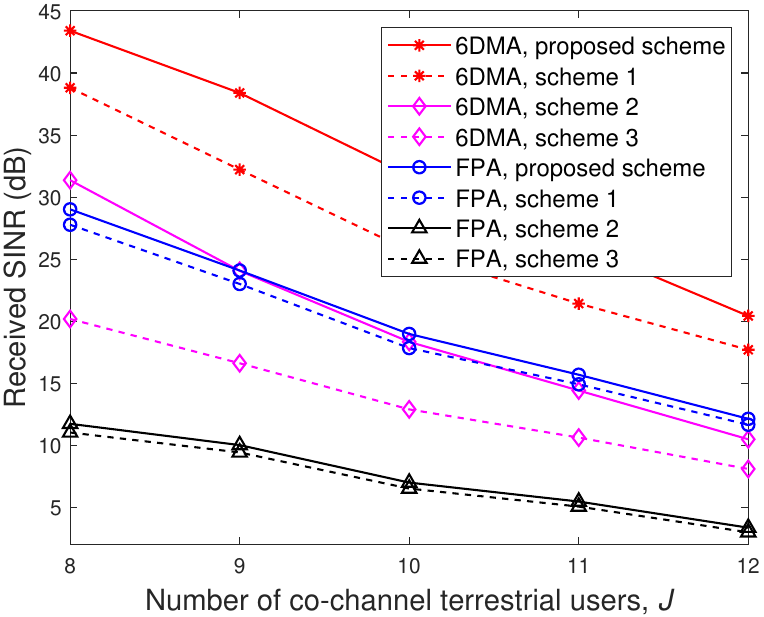}
		\caption{UAV-received SINR versus the number of co-channel users.}
		\label{fig:result_J}
	\end{minipage}
	\begin{minipage}{0.32\linewidth}
		\centering
		\includegraphics[width=0.85\linewidth]{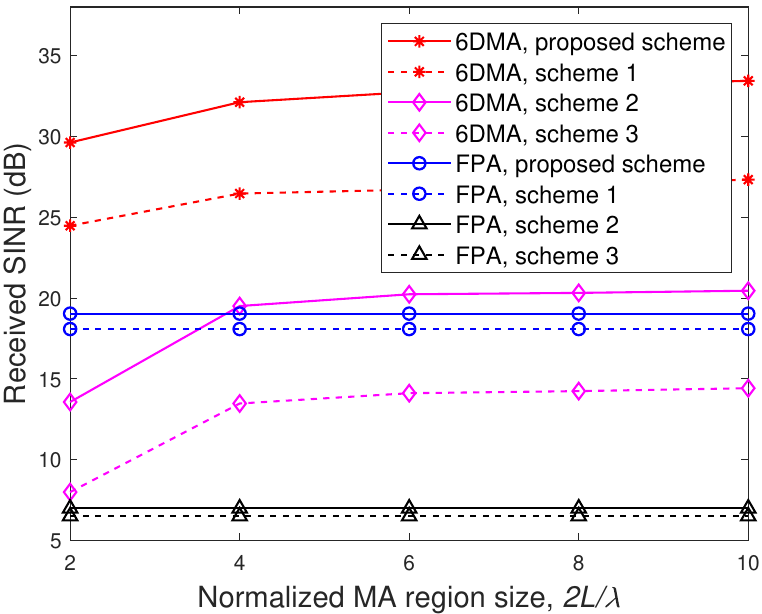}
		\caption{UAV-received SINR versus the normalized MA region size.}
		\label{fig:result_L}
	\end{minipage}
         \begin{minipage}{0.32\linewidth}
		\centering
		\includegraphics[width=0.85\linewidth]{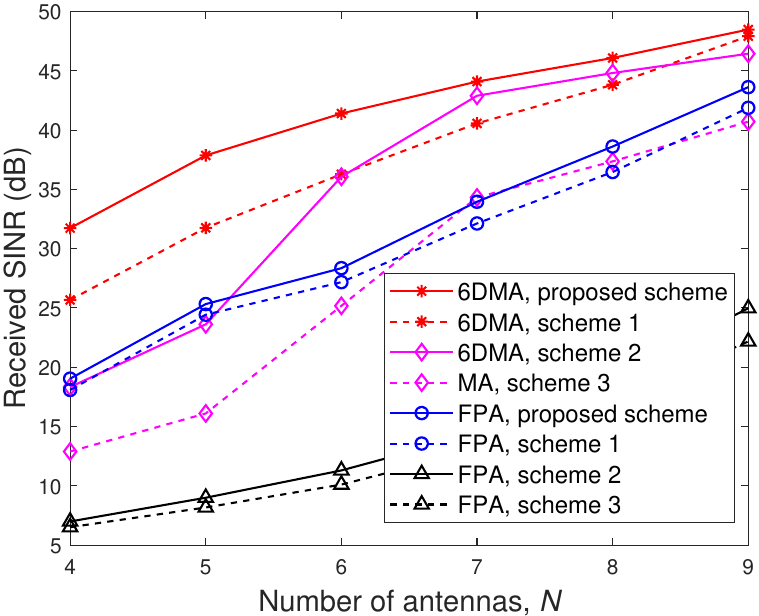}
		\caption{UAV-received SINR versus the number of antennas.}
		\label{fig:result_N}
	\end{minipage}
        \vspace{-0.4cm} 
\end{figure*}

First, we evaluate the UAV-received SINR versus the number of co-channel terrestrial users $J$ (i.e., the number of occupied BSs). The normalized MA region size is set as $2L/\lambda=4$, and the number of antennas is set as $N=4$. It is observed from Fig.~\ref{fig:result_J} that the UAV-received SINR of all schemes based on MA and FPA decreases as $J$ increases. The schemes based on MA always outperform those based on FPA, and our proposed scheme outperforms all three baseline schemes. Additionally, compared to FPA-based schemes, the proposed scheme with MA array has $15$~dB SINR improvement even with $J=12$ co-channel users, showing the superiority of the proposed method for interference mitigation under a large number of interference BSs.

Then, we explore the UAV-received SINR versus the normalized MA region size $2L/\lambda$, while fixing the number of co-channel terrestrial users as $J=10$ and the number of antennas as $N=4$. As shown in Fig.~\ref{fig:result_L}, the received SINR of the MA-based schemes increases with $2L/\lambda$, whereas that of the FPA-based schemes remains constant. This is because a larger region size provides more DoFs in antenna position optimization, thereby enhancing interference mitigation performance.

Finally, Fig.~\ref{fig:result_N} illustrates the UAV-received SINR versus the number of antennas $N$, where the normalized MA region size is fixed as $2L/\lambda=4$, and the number of co-channel terrestrial users is fixed as $J=10$. It is observed that the received SINR increases with $N$ for all schemes, and the MA-based schemes always outperform the FPA-based ones. Furthermore, as the number of antennas increases, the performance gain provided by array rotation decreases. This result shows that when the number of antennas is sufficiently large, antenna position optimization offers a more significant performance enhancement for cellular-connected UAV interference mitigation than the array rotation.

\section{Conclusion}\label{sec:con}

In this letter, we investigated the utilization of 6DMA array to enhance the interference mitigation in cellular-connected UAV communication systems. In particular, we developed a BCD-based algorithm to solve the SINR maximization problem by jointly optimizing the APV, ARV, receive beamforming vector, and selection of the associated BS. Numerical results demonstrated that the 6DMA-based schemes outperform those with the traditional FPA arrays and significantly enhance the received SINR of the UAV. 

\bibliographystyle{IEEEtran}

\bibliography{MA_UAV}

\end{document}